\documentstyle[aps]{revtex}
\begin{document}
\voffset 0.8truecm
\title{Additivity 
of entanglement of formation for some special cases
}

\author{
Heng Fan
}
\address{
Quantum computation and quantum information project,
ERATO, \\
Japan Science and Technology Corporation,\\
Daini Hongo White Building.201,
Hongo 5-28-3, Bunkyo-ku, Tokyo 133-0033, Japan.
}
\maketitle

\begin{abstract}
The proof of additivity
of entanglement of formation for some special cases is given.
The strong concavity of von Neumann entropy due to
strong subadditivity of von Neumann entropy is presented.
Some general relations concerning about the entanglement
of formation are proposed.
\end{abstract}
                                                 

\section{Introduction and some general results}
Entanglement of formation (EoF) is a widely accepted measurement of
entanglement\cite{BDSW}. For a bipartite state $\rho _{AB}$ in
Hilbert space $H_A\otimes H_B$,
the entanglement of
formation is defined as
\begin{eqnarray}
E_f(\rho _{AB})=
min\sum _i{p_i}
S\left( {\rm Tr}_A|\Psi _{AB}^i\rangle \langle \Psi _{AB}^i|\right),
\end{eqnarray}
where the minimization is over all possible ensembles such that
$\rho _{AB}=\sum _ip_i|\Psi _{AB}^i\rangle \langle \Psi _{AB}^i|$,
$S(\rho )=-{\rm Tr}\rho \log _2\rho $ is the von Neumann entropy.

It is a long-standing conjecture that for product density matrix
$\rho _{AB}\otimes \rho _{A'B'}$ in Hilbert space
$H_A\otimes H_B\otimes H_{A'}\otimes H_{B'}$
the EoF is additive:
\begin{eqnarray}
E_f(\rho _{AB}\otimes \rho _{A'B'})
=E_f(\rho _{AB})+E_f(\rho _{A'B'}).
\end{eqnarray}
But only a few cases are proved.

From the definition we know that EoF
is weak additive, i.e., the following inequality holds:
\begin{eqnarray}
E_f(\rho _{AB}\otimes \rho _{A'B'})
\le E_f(\rho _{AB})+E_f(\rho _{A'B'}).
\end{eqnarray}
In order to prove the additivity of EoF,
we just need to prove the opposite inequality. 

First, we present some relations 
concerning about the EoF, trivial or not:
\begin{eqnarray}
min\sum _ip_i\left( S(\rho _A^i)+S(\rho _{A'}^i)\right)
&=&
min\sum _ip_i\left( S(\rho _A^i) +E_f(\rho _{A'B'}^i)\right) 
\label{EoF1}
\\
&=&
min\sum _ip_i\left( E_f(\rho _{AB}^i)+S(\rho _{A'}^i)\right) 
\label{EoF2}
\\
&=&
min\sum _ip_i\left( E_f(\rho _{AB}^i) +E_f(\rho _{A'B'}^i)\right) 
\label{EoF3}
\\
&=&E_f(\rho _{AB})+E_f(\rho _{A'B'}),
\label{EoF4}
\end{eqnarray}
where we have pure state decomposition
\begin{eqnarray}
\rho _{AB}\otimes \rho _{A'B'}=\sum _ip_i|\Psi ^i_{ABA'B'}
\rangle \langle \Psi ^i_{ABA'B'}|.
\label{decomp0}
\end{eqnarray}
Here we denote
$\rho ^i_{A}=Tr_{BA'B'}
(|\Psi ^i_{ABA'B'}
\rangle \langle \Psi ^i_{ABA'B'}|)$, and similar notations
are also used for other states with partial traces.
We remark that optimal pure state decomposition in one
mininization is not necessarily the optimal decomposition in another
mininization. The proof of these relations are straightforward.

We can find from these relations that in order to prove
the additivity of EoF, we need to prove
the EoF
\begin{eqnarray}
E_f(\rho _{AB}\otimes \rho _{A'B'})=\sum _ip_iS(\rho _{AA'}^i),
\label{whole}
\end{eqnarray}
is at least one quantities appeared in the minization in
relations (\ref{EoF1},\ref{EoF2},\ref{EoF3},\ref{EoF4}),
where we suppose (\ref{decomp0}) is optimal pure states
decomposition for EoF.
However, it seems unlikely to use directly the first quantities since
we have subadditivity inequality,
\begin{eqnarray}
S(\rho _{AA'}^i)\le S(\rho _{A}^i)+S(\rho _{A'}^i),
\end{eqnarray}
where the equality holds if and only if
$\rho _{AA'}^i=\rho _{A}^i\otimes \rho _{A'}^i$ which can not
be satisfied in general in (\ref{whole}). 
In this paper, we will use the second or the third
quantities in (\ref{EoF1}, \ref{EoF2}) to
prove the additivity of EoF for some special cases. 

\section{Strong concavity of von Neumann entropy}
Before we proceed, we give some useful relations\cite{NC},
\begin{eqnarray}
S(\sum _ip_i\rho ^i\otimes |i\rangle \langle i|)
=H(p_i)+\sum _ip_iS(\rho ^i).
\label{useful}
\end{eqnarray}
where $H(p_i)$ is Shannon entropy.
We also have the strong concavity of von Neumann entropy
\begin{eqnarray}
S(\sum _ip_i\rho _1^i\otimes \rho _2^i)
\ge \sum _ip_iS(\rho _1^i)+S(\sum _ip_i\rho _2^i), 
\label{strong1}
\\
S(\sum _ip_i\rho _1^i\otimes \rho _2^i)
\ge S(\sum p_i\rho _1^i)+\sum _ip_iS(\rho _2^i), 
\label{strong2}
\end{eqnarray}
due to the strong subadditivity of von Neumann entropy\cite{R}
\begin{eqnarray}
S(\rho _{123})+S(\rho _2)
\le S(\rho _{12})+S(\rho _{23}). 
\label{subadd}
\end{eqnarray}
We define $\rho _{123}=\sum _ip_i\rho _1^i\otimes \rho _2^i\otimes
|i\rangle _3\langle i|$, and use the relation (\ref{useful}),
and we have
\begin{eqnarray}
S(\rho _{12})&\ge &
S(\rho _{123})+S(\rho _2)-S(\rho _{23})\nonumber \\
&=&H(p_i)+\sum _ip_iS(\rho _1^i\otimes \rho _2^i)
+S(\sum _ip_i\rho _2^i)-H(p_i)-\sum _ip_iS(\rho _2^i)
\nonumber \\
&=&\sum _ip_iS(\rho _1^i)+S(\sum_ip_i\rho _2^i).
\end{eqnarray}
So, we proved the relation (\ref{strong1}),similarly for (\ref{strong2}). 
Relations (\ref{strong1},\ref{strong2}) is stronger than the
relation due to concavity of
von Neumann entropy,
\begin{eqnarray}
S(\sum _ip_i\rho _1^i\otimes \rho _2^i)
\ge \sum _ip_i\left( S(\rho _1^i)+S(\rho _2^i)\right) .
\end{eqnarray}
Comparing (\ref{strong1}) and (\ref{useful}), we know
\begin{eqnarray}
H(p_i)\ge S(\sum _ip_i\rho _i).
\end{eqnarray}
This was proved in ref.\cite{We}, and reproved by a different
method in ref.\cite{HJW}, here we provide a different proof.
The strong subadditivity of von Neumann entropy (\ref{subadd})
was recently used to prove the additivity of eantanglement
breaking channel\cite{S} and entanglement cost\cite{V} where
essentially the strong concavity of von Neumann entropy
(\ref{strong1},\ref{strong2}) were used.
Since the relations (\ref{strong1},\ref{strong2}) are very useful
and deserve an independent name,
we call them
the strong concavity of von Neumann entropy in this paper and
use them directly without tracking back to the strong subadditivity
of von Neumann entropy.

\section{Additivity of EoF for special case I}
Next we consider a special class of pure states,
\begin{eqnarray}
|\Psi _{ABA'B'}\rangle
=\sum _{\alpha \beta }
\sqrt {\lambda _{\alpha \beta }}|\alpha \rangle _A
|\alpha \rangle _B|\beta \rangle _{A'}|\beta \rangle _{B'},
\label{special}
\end{eqnarray}
we know
\begin{eqnarray}
\rho _{AB}&=&Tr_{A'B'}
(|\Psi _{ABA'B'}\rangle
\langle \Psi _{ABA'B'}|)
\nonumber \\
&=&\sum _{\beta }\left(
\sum _{\alpha }\sqrt{\lambda _{\alpha \beta }}
|\alpha \rangle _A|\alpha \rangle _B\right)
\left(
\sum _{\alpha '}\sqrt{\lambda _{\alpha '\beta }}_A
\langle \alpha '|_B\langle \alpha '|\right)
\nonumber \\
&=&\sum _{\beta }\lambda _{\beta }
|\Psi ^{\beta }_{AB}\rangle
\langle \Psi ^{\beta }_{AB}|,
\end{eqnarray}
where we use the notation
\begin{eqnarray}
|\Psi ^{\beta }_{AB}\rangle
\equiv \frac {1}{\sqrt{\lambda _{\beta }}}\left( \sum _{\alpha }
\sqrt{\lambda _{\alpha \beta }}
|\alpha \rangle _A|\alpha \rangle _B\right) ,
~~~\lambda _{\beta }\equiv \sum _{\alpha }\lambda _{\alpha \beta }. 
\end{eqnarray}
Similarly, we have the result for $\rho _{A'B'}$ 
\begin{eqnarray}
\rho _{A'B'}&=&
\sum _{\alpha }\left(
\sum _{\beta }\sqrt{\lambda _{\alpha \beta }}
|\beta \rangle _{A'}|\beta \rangle _{B'}\right)
\left(
\sum _{\beta '}\sqrt{\lambda _{\alpha \beta '}}_{A'}
\langle \beta '|_{B'}\langle \beta '|\right)
\nonumber \\
&=&\sum _{\alpha } \lambda _{\alpha }
|\Psi ^{\alpha }_{A'B'}\rangle
\langle \Psi ^{\alpha }_{A'B'}|,
\label{decomposition}
\end{eqnarray}
and we use the notations
\begin{eqnarray}
|\Psi ^{\alpha }_{A'B'}\rangle
\equiv \frac {1}{\sqrt{\lambda _{\alpha }}}\left( \sum _{\beta }
\sqrt{\lambda _{\alpha \beta }}
|\beta \rangle _{A'}|\beta \rangle _{B'}\right) ,
~~~\lambda _{\alpha  }\equiv \sum _{\beta }\lambda _{\alpha \beta }. 
\end{eqnarray}
We also know the reduced density operator
\begin{eqnarray}
\rho _{AA'}&=&\sum _{\alpha \beta }
\lambda _{\alpha \beta }|\alpha \rangle _A\langle \alpha |
\otimes |\beta \rangle _{A'}\langle \beta |
\nonumber \\
&=&\sum _{\alpha }\lambda _{\alpha }
|\alpha \rangle _A\langle \alpha |\otimes
Tr_{B'}(|\Psi ^{\alpha }_{A'B'}\rangle
\langle \Psi ^{\alpha }_{A'B'}). 
\end{eqnarray}
Then by using the relation (\ref{useful}), we obtain
\begin{eqnarray}
S(\rho _{AA'})&=&H(\lambda _{\alpha })
+\sum _{\alpha } \lambda _{\alpha }S\left( Tr_{B'}
(|\Psi ^{\alpha }_{A'B'}\rangle
\langle \Psi ^{\alpha }_{A'B'}|)\right)
\nonumber \\
&=&S(\rho _A)+\sum _{\alpha }\lambda _{\alpha }
S\left( Tr_{B'}
(|\Psi ^{\alpha }_{A'B'}\rangle
\langle \Psi ^{\alpha }_{A'B'}|)\right) .
\label{medium}
\end{eqnarray}
Due to pure state decomposition of $\rho _{A'B'}$ presented
in (\ref{decomposition}),
we have
\begin{eqnarray}
\sum _{\alpha }\lambda _{\alpha }
S\left( Tr_{B'}
(|\Psi ^{\alpha }_{A'B'}\rangle
\langle \Psi ^{\alpha }_{A'B'}|)\right) \ge
E_f(\rho _{A'B'}).
\end{eqnarray}
Thus 
\begin{eqnarray}
S(\rho _{AA'})\ge S(\rho _A)
+E_f(\rho _{A'B'}).
\label{skew}
\end{eqnarray}
Similar to relation (\ref{medium}), we have
\begin{eqnarray}
S(\rho _{AA'})&=&
S(\rho _{A'})+\sum _{\beta }\lambda _{\beta }
S\left( Tr_{B}
(|\Psi ^{\beta }_{AB}\rangle
\langle \Psi ^{\beta }_{AB}|)\right) .
\label{medium2}
\end{eqnarray}
We conclude the superadditivity of EoF
is ture for pure state $|\Psi _{ABA'B'}\rangle $.
The relations (\ref{medium}) and (\ref{skew}) are even stronger than the
superadditivity of entanglement of formation. We summarize
\begin{eqnarray}
S(\rho _{AA'})\ge E_f(\rho _{AB})
+E_f(\rho _{A'B'}).
\end{eqnarray}

In case all pure states in the decomposition of
$\rho _{AB}\otimes \rho _{A'B'}$ can be written as
the form (\ref{special}) by independent Schmidt decomposition
on $H_A\otimes H_B$ and $H_{A'}\otimes H_{B'}$, the
additivity of EoF holds since we have
\begin{eqnarray}
E_f(\rho _{AB}\otimes \rho _{A'B'})
&=&\sum _ip_iE_f(|\Psi _{ABA'B'}^i\rangle )
\nonumber \\
&\ge &\sum _ip_i\left( E_f(\rho _{AB}^i)+E_f(\rho _{A'B'}^i)\right) 
\nonumber \\
&\ge &E_f(\rho _{AB})+E_f(\rho _{A'B'}),
\end{eqnarray}
where we assume (\ref{decomp0}) is the optimal decomposition for EoF.
We remark that the additivity of EoF \cite{MSW,V}
deduced from additivity
of entanglement breaking channel\cite{S} belong to
the class (\ref{special}).

For an arbitrary pure state $|\tilde{\Psi }_{ABA'B'}\rangle $, we can
always find unitary transformations $U_{AA'}$ and
$V_{BB'}$ in $H_A\otimes H_{A'}$ and $H_B\otimes H_{B'}$ respectively,
and transfer $|\tilde{\Psi }_{ABA'B'}\rangle $ to (\ref{special}),
\begin{eqnarray}
U_{AA'}\otimes V_{BB'}|\tilde{\Psi }_{ABA'B'}\rangle
&=&|\Psi _{ABA'B'}\rangle 
\nonumber \\
&=&\sum _{\alpha \beta }
\sqrt {\lambda _{\alpha \beta }}|\alpha \rangle _A
|\alpha \rangle _B|\beta \rangle _{A'}|\beta \rangle _{B'}.
\end{eqnarray}
We know
\begin{eqnarray}
S(\tilde {\rho }_{AA'})&=&S(\rho _{AA'})
\nonumber \\
&=&S(\rho _A)+\sum _{\alpha }\lambda _{\alpha }
S\left( Tr_{B'}
(|\Psi ^{\alpha }_{A'B'}\rangle
\langle \Psi ^{\alpha }_{A'B'}|)\right) ,
\end{eqnarray}
but it is not clear whether in general the relation
\begin{eqnarray}
S(\tilde {\rho }_{AA'})\ge E_f(\tilde {\rho }_{AB})
+E_f(\tilde {\rho }_{A'B'})
\end{eqnarray}
holds or not, this relation is called superadditivity
of EoF\cite{VW,MSW}. If the superadditivity
of EoF holds for arbitrary pure states, the additivity
of EoF follows directly.

\section{Additivity of EoF for special case II}
In what follows, we restate the HJW theorem presented in ref.\cite{HJW}:
For any density matrix $\rho $ having the diagonal form
\begin{eqnarray}
\rho =\sum _i\lambda _i|e_i\rangle \langle e_i|,
\end{eqnarray}
can be written as the mixed states of $|\psi _i\rangle $ with
probability $p_i$
\begin{eqnarray}
\rho =\sum _ip_i|\psi _i\rangle \langle \psi _i|,
\end{eqnarray}
iff there exists a unitary transformation $U$ such that
\begin{eqnarray}
|\psi _i\rangle =\frac {1}{\sqrt{p_i}}\sum _jU_{ij}
\sqrt{\lambda _j}|e_j\rangle .
\end{eqnarray}     

Now we present another class of states for which
the additivity of EoF holds.
Consider about two density matrices in $H_A\otimes H_B$ and
$H_{A'}\otimes H_{B'}$,
\begin{eqnarray}
\rho _{AB}=\sum _J\lambda _J|J\rangle _{AB}\langle J|,\\
\rho _{A'B'}=\sum _K\lambda _K|K\rangle _{A'B'}\langle K|,
\end{eqnarray}
where $\lambda _J$ and $|J\rangle _{AB}$ are eigenvalues
and eigenvectors of density operator $\rho _{AB}$,
and similarly for $\rho _{A'B'}$.

Let's assume that the eigenvectors $|J\rangle _{AB}$ have
some special properties. Suppose
$|J\rangle _{AB}=
\sum a_{j_1j_2}|j_1\rangle _A|j_2\rangle _B$,
$|J'\rangle _{AB}=
\sum a_{j'_1j'_2}|j'_1\rangle _A|j'_2\rangle _B$,
$J\not =J'$,
we assume $|j_1\rangle _A\not=|j_1'\rangle _A$,
$|j_2\rangle _B\not=|j_2'\rangle _B$.
For example, $\rho _{AB}=\lambda |00\rangle \langle 00|
+(1-\lambda ){1\over 2}(|11\rangle +|22\rangle )
(\langle 11|+\langle 22|)$. We assume $|K\rangle _{A'B'}$
also has this property.

The Schmidt decomposition of states $|J\rangle _{AB}$ and
$|K\rangle _{A'B'}$ have the following form:
\begin{eqnarray}
|J\rangle _{AB}=\sum _{\alpha _J}\sqrt{\eta _{\alpha _J}}
|\alpha _J\rangle _A|\alpha _J\rangle _B, \\
|K\rangle _{A'B'}=\sum _{\beta _K}\sqrt{\xi _{\beta _K}}
|\beta _K\rangle _{A'}|\beta _K\rangle _{B'},
\end{eqnarray}
Since $|J\rangle _{AB}$ and  $|J'\rangle _{AB}$ have
the property presented above,
we have the following relation
\begin{eqnarray}
\langle \alpha _J|\alpha '_{J'}\rangle =\delta _{J,J'}
\delta _{\alpha _J,\alpha '_J}.
\label{delta1}
\end{eqnarray}
This can be understood that $\rho _{AB}$ is block diagonal, so
besides $\langle \alpha _J|\alpha '_{J}\rangle
=\delta _{\alpha _J,\alpha '_J}$ inside one block, we also
have orthogonal relation for different blocks.
Similarly, we have
\begin{eqnarray}
\langle \beta _K|\beta '_{K'}\rangle =\delta _{K,K'}
\delta _{\beta _K,\beta '_K}.
\label{delta2}
\end{eqnarray}

Now, let's consider the pure states decomposition,
\begin{eqnarray}
\rho _{AB}\otimes \rho _{A'B'}
=\sum _ip_i|\Psi ^i_{ABA'B'}\rangle \langle \Psi ^i_{ABA'B'}|.
\label{decomp}
\end{eqnarray}
Due to HJW theorem \cite{HJW}
we can find a unitary matrix $U_{i,JK}$ and also
considering about the Schmidt decomposition, we have
\begin{eqnarray}
|\Psi ^i_{ABA'B'}\rangle &=&\frac {1}{\sqrt{p_i}}
\sum _{JK}U_{i,JK}\sqrt{\lambda _J\lambda _K}
|J\rangle _{AB}\otimes |K\rangle _{A'B'},\\
&=&\frac {1}{\sqrt{p_i}}
\sum _{JK}U_{i,JK}\sqrt{\lambda _J\lambda _K}
(\sum _{\alpha _J}\sqrt{\eta _{\alpha _J}}
|\alpha _J\rangle _A |\alpha _J\rangle _B)\otimes
(\sum _{\beta _K}\sqrt{\xi _{\beta _K}}
|\beta _K\rangle _{A'} |\beta _K\rangle _{B'}).
\end{eqnarray}
We remark that the properties of eigenvectors already used here.
So, the reduced density operator in $H_A\otimes H_{A'}$ can be
obtained as
\begin{eqnarray}
\rho _{AA'}^i&=&Tr_{BB'}\left( |\Psi ^i_{ABA'B'}\rangle
\langle \Psi ^i_{ABA'B'}|\right)
\nonumber \\
&=&\frac {1}{p_i}
\sum _{JKJ'K'}\sum _{\alpha _J\alpha '_{J'}\beta _K\beta '_{K'}}
U_{i,JK}U^*_{i,J'K'}\sqrt{\lambda _J\lambda _{K}
\lambda _{J'}\lambda _{K'}}\sqrt{\eta _{\alpha _J}
\eta _{\alpha '_{J'}}\xi _{\beta _K}\xi _{\beta '_{K'}}}
\nonumber \\
&&|\alpha _J\rangle _A\langle \alpha '_{J'}|\otimes
|\beta _K\rangle _{A'}\langle \beta '_{K'}|\delta _{J,J'}
\delta _{\alpha _J,\alpha '_J}\delta _{K,K'}
\delta _{\beta _K,\beta '_K}\\
&=&\frac {1}{p_i}
\sum _{JK}\sum _{\alpha _J\beta _K}
U_{i,JK}U^*_{i,JK}\lambda _J\lambda _{K}\eta _{\alpha _J}
\xi _{\beta _K}
|\alpha _J\rangle _A\langle \alpha _{J}|\otimes
|\beta _K\rangle _{A'}\langle \beta _{K}|,
\end{eqnarray}
where we have used the relations (\ref{delta1},\ref{delta2}) when
we take trace in $H_B\otimes H_{B'}$.
With the help of strong cancavity relation (\ref{strong1}),
the von Neumann entropy of $\rho ^i_{AA'}$ has the following form,
\begin{eqnarray}
S(\rho ^i_{AA'})&=&
S\left( \sum _K\lambda _K[\frac {1}{p_i}
\sum _JU_{i,JK}U^*_{i,JK}\lambda _J\sum _{\alpha _J}\eta _{\alpha _J}
|\alpha _J\rangle _A\langle \alpha _J|]\otimes
[\sum _{\beta _K}\xi _{\beta _K}
|\beta _K\rangle _{A'}\langle \beta _{K}|]\right)
\nonumber \\
&\ge &\sum _K\lambda _K
S\left( \frac {1}{p_i}
\sum _JU_{i,JK}U^*_{i,JK}\lambda _J\sum _{\alpha _J}\eta _{\alpha _J}
|\alpha _J\rangle _A\langle \alpha _J|\right) 
+S\left( \sum _K\lambda _K
\sum _{\beta _K}\xi _{\beta _K}
|\beta _K\rangle _{A'}\langle \beta _{K}|\right)
\nonumber \\
&=&\sum _K\lambda _K
S\left( \rho _{A}^{iK}\right) 
+S\left( \rho _{A'}^i\right),
\label{key}
\end{eqnarray}
where we used the the following notations and relations
\begin{eqnarray}
\rho _{A}^{iK}
&\equiv &
\frac {1}{p_i}
\sum _JU_{i,JK}U^*_{i,JK}\lambda _J\sum _{\alpha _J}\eta _{\alpha _J}
|\alpha _J\rangle _A\langle \alpha _J|, 
\\
\rho _{A'}^i
&=&\sum _K\lambda _K\sum _{\beta _K}\xi _{\beta _K}
|\beta _K\rangle _{A'}\langle \beta _{K}|\\
&=&Tr_A(\rho _{AA'}^i).
\end{eqnarray}
In the same time, we can identify
\begin{eqnarray}
\rho _{A}^{iK}
&=&
Tr_{B}
\frac {1}{p_i}
\sum _{JJ'}U_{i,JK}U^*_{i,J'K}
\sqrt{\lambda _J\lambda _{J'}}
\sum _{\alpha _J\alpha '_{J'}}
\sqrt{\eta _{\alpha _J}\eta _{\alpha '_{J'}}}
|\alpha _J\rangle _A\langle \alpha '_{J'}|
\otimes
|\alpha _J\rangle _{B}\langle \alpha '_{J'}|
\nonumber \\        
&=&Tr_{B}|\Psi _{AB}^{iK}\rangle \langle \Psi _{AB}^{iK}|,
\end{eqnarray}
where, we use the definition
\begin{eqnarray}
|\Psi _{AB}^{iK}\rangle \equiv
\frac {1}{\sqrt{p_i}}
\sum _{J}U_{i,JK}
\sqrt{\lambda _J}
\sum _{\alpha _J}
\sqrt{\eta _{\alpha _J}}|\alpha _J\rangle _A
|\alpha _J\rangle _{B}.
\end{eqnarray}
We can find the reduced density operator
$\rho _{AB}^i=Tr_{A'B'}(|\Psi ^i_{ABA'B'}\rangle
\langle \Psi _{ABA'B'}|)$ has the pure state decomposition
\begin{eqnarray}
\rho _{AB}^i=\sum _K\lambda _K
|\Psi _{AB}^{iK}\rangle \langle \Psi _{AB}^{iK}|.
\end{eqnarray}
Thus we obtain the result from relation (\ref{key}),
\begin{eqnarray}
E_f(|\Psi ^i_{ABA'B'}\rangle )&=&S(\rho _{AA'}^i)
\nonumber \\
&\ge &
E_f(\rho _{AB}^i)+E_f(\rho _{A'B'}^i).
\end{eqnarray}
Here we use the fact
\begin{eqnarray}
E_f(\rho _{AB}^i)&=&E_f(\sum _K\lambda _K
|\Psi _{AB}^{iK}\rangle \langle \Psi _{AB}^{iK}|)
\le \sum _K\lambda _KS(\rho _{A}^{iK}),\\
E_f(\rho _{A'B'}^i)&\le &S(\rho _{A'}^i).
\end{eqnarray}
Here we summarize that the EoF of
pure state $|\Psi ^i_{ABA'B'}\rangle $ is at least
the sum of entanglement of formation of
$\rho _{AB}^i$ and $\rho _{A'B'}^i$ provided the pure state
decomposition relation (\ref{decomp}),
\begin{eqnarray}
E_f(|\Psi _{ABA'B'}^i\rangle )\ge E_f(\rho ^i_{AB})
+E_f(\rho ^i_{A'B'}).
\label{super}
\end{eqnarray}

What follows seems straightforward.
Suppose (\ref{decomp}) is the optimal decomposition for
EoF, we have
\begin{eqnarray}
E_f(\rho _{AB}\otimes \rho _{A'B'})
&=&\sum _ip_iE_f(|\Psi _{ABA'B'}^i\rangle )
\nonumber \\
&\ge &\sum _ip_i\left( E_f(\rho _{AB}^i)+E_f(\rho _{A'B'}^i)\right) 
\nonumber \\
&\ge &E_f(\rho _{AB})+E_f(\rho _{A'B'}),
\end{eqnarray}
the last inequality is due to the fact
\begin{eqnarray}
\rho _{AB}=\sum _ip_i\rho _{AB}^i, ~~~
\rho _{A'B'}=\sum _ip_i\rho _{A'B'}^i.
\end{eqnarray}
We already know the EoF of
$\rho _{AB}\otimes \rho _{A'B'}$ is at most the
sum of entanglement of formation of $\rho _{AB}$ and $\rho _{A'B'}$
\begin{eqnarray}
E_f(\rho _{AB}\otimes \rho _{A'B'})
\le E_f(\rho _{AB})+E_f(\rho _{A'B'}),
\end{eqnarray}
thus
we have the additivity of EoF for this special case.
\begin{eqnarray}
E_f(\rho _{AB}\otimes \rho _{A'B'})
=E_f(\rho _{AB})+E_f(\rho _{A'B'}).
\end{eqnarray}

\section{Summary and discussion}
Though we present here the additivity of EoF for some special
cases, the problem of additivity of EoF for
general case is still open.

For pure state in the decomposition,
\begin{eqnarray}
|\Psi ^i_{ABA'B'}\rangle &=&\frac {1}{\sqrt{p_i}}
\sum _{JK}U_{i,JK}\sqrt{\lambda _J\lambda _K}
|J\rangle _{AB}\otimes |K\rangle _{A'B'},
\end{eqnarray}
we denote
\begin{eqnarray}
|\Psi _{AB}^{iK}\rangle &\equiv &
\frac {1}{\sqrt{p_i}}
\sum _{J}U_{i,JK}\sqrt{\lambda _J}
|J\rangle _{AB},
~~~\rho _{A}^{iK}\equiv Tr_B|\Psi _{AB}^{iK}\rangle \langle
\Psi _{AB}^{iK}|
;
\\
|\Psi _{A'B'}^{iJ}\rangle &\equiv &
\frac {1}{\sqrt{p_i}}
\sum _{K}U_{i,JK}\sqrt{\lambda _K}
|K\rangle _{A'B'},
~~~\rho _{A'}^{iJ}\equiv Tr_{B'}
|\Psi _{A'B'}^{iJ}\rangle \langle
\Psi _{A'B'}^{iJ}|;
\\
\rho _{AA'}^i&\equiv &
Tr_{BB'}|\Psi ^i_{ABA'B'}\rangle \langle
\Psi ^i_{ABA'B'}|.
\end{eqnarray}
Here, similar as the superadditivity of EoF, we
may ask the question: whether
the following relation holds:
\begin{eqnarray}
S(\rho _{AA'}^i)\ge
\sum _K\lambda _KS(\rho _{A}^{iK})
+\sum _J\lambda _JS(\rho _{A'}^{iJ})?
\label{question1}
\end{eqnarray}
One possible method to answer this question is to know whether
the following relation is true:
\begin{eqnarray}
S(\rho _{AA'}^i)
&\ge &
S\left( \sum _K\lambda _K\rho _{A}^{iK}\otimes
(Tr_{B'}|K\rangle _{A'B'}\langle K|)\right)
+
S\left( \sum _J\lambda _J
(Tr_{B}|J\rangle _{AB}\langle J|)\otimes
\rho _{A'}^{iJ}\right)
\nonumber \\
&&
-S\left( \sum _{JK}\frac {1}{p_i}\lambda _J\lambda _K|U_{i,JK}|^2
(Tr_{B}|J\rangle _{AB}\langle J|)\otimes
(Tr_{B'}|K\rangle _{A'B'}\langle K|)\right) ?
\label{question2}
\end{eqnarray}
We can obtain (\ref{question1}) from (\ref{question2}) by using
strong concavity of von Neumann entropy for the first two terms
and the subadditivity for the third term.

We summarize our idea here: a, relation (\ref{question2}) is a sufficient
condition of relation (\ref{question1});
b, relation (\ref{question1}) is a sufficient condition for
the additivity of EoF. But both a or b may not be necessary conditions. 

Werner state\cite{Werner} is an interesting state to check
the superadditivity of the EoF since the EoF of Werner state
is known\cite{VW}. Suppose we have an Werner state in
$H_{A}\otimes H_{B}\otimes H_{A'}\otimes H_{B'}$ with
dimension 2 for each Hilbert space. The reduced
density operators in $H_{A}\otimes H_{B}$ and
$H_{A'}\otimes H_{B'}$ are still Werner states, and
EoF are known. The result shows that the
superadditivity of the EoF is correct for this case.

{\it Acknowlegements}: The author would like
to thank K.Matsumoto for
pointing out a mistake when the author begin to study
this problem, A.Winter for introducing the superadditivity of
entanglement of formation,  R.F.Werner for pointing out
the superadditivity was first proposed in their paper and
encouraging the author to continue to study this problem when
the group in Braunschweig (K.G.H.Vollbrecht and M.Wolf)
found a mistake in the author's previous results which was
also found by the author himself.
The author also would like to thank members in ERATO project,
M.Hamada, T.Shimono, A.Miyake, Y.Tsuda, X.B.Wang, F.Yura et al
for useful discussions and H.Imai for support.

\end{document}